\documentclass[12pt,preprint]{aastex}
\usepackage{float,epsfig,psfig}

\newbox\grsign \setbox\grsign=\hbox{$>$} 
\newdimen\grdimen \grdimen=\ht\grsign
\newbox\laxbox \newbox\gaxbox
\setbox\gaxbox=\hbox{\raise.5ex\hbox{$>$}\llap
     {\lower.5ex\hbox{$\sim$}}}\ht1=\grdimen\dp1=0pt
\setbox\laxbox=\hbox{\raise.5ex\hbox{$<$}\llap
     {\lower.5ex\hbox{$\sim$}}}\ht2=\grdimen\dp2=0pt

\newcommand{\cha}{{\sl Chandra}}
\newcommand{\ein}{{\sl Einstein}}

\newcommand{\xmm}{XMM-{\sl Newton}}
\newcommand{\sax}{{\sl Beppo}-SAX}
\newcommand{\xte}{{\sl Rossi}-XTE}

\newcommand{\nsh}{$10^{21}\; {\rm cm^{-2}}$}
\newcommand{\kev}{${\rm ke\!V}$}

\begin{document}

\title{
{\sl Chandra} Phase-Resolved X-ray Spectroscopy of the Crab Pulsar}

\author{
Martin C. Weisskopf \altaffilmark{1},
Stephen L. O'Dell \altaffilmark{1},
Frits Paerels \altaffilmark{2},
Ronald F. Elsner \altaffilmark{1},
Werner Becker \altaffilmark{3},
Allyn F. Tennant \altaffilmark{1}, and
Douglas A. Swartz \altaffilmark{4}
}
\altaffiltext{1}
{Space Science Department, NASA Marshall Space Flight Center, SD50, Huntsville,
AL 35812}
\altaffiltext{2}
{Columbia Astrophysics Laboratory and Physics Department, Columbia University,
New York, NY 10027}
\altaffiltext{3}
{Max Planck Institut f\"ur Extraterrestrische Physik, 85740 Garching bei
M\"unchen, Germany}
\altaffiltext{4}
{Universities Space Research Association, NASA Marshall Space Flight Center,
SD50, Huntsville, AL 35812}

\begin{abstract}
We present the first phase-resolved study of the X-ray spectral properties of
the Crab Pulsar that covers all pulse phases.
The superb angular resolution of the \cha\ X-ray Observatory enables
distinguishing the pulsar from the surrounding nebulosity, even at pulse
minimum.
Analysis of the pulse-averaged spectrum measures interstellar X-ray extinction
due primarily to photoelectric absorption and secondarily to scattering by dust
grains in the direction of the Crab Nebula.
We confirm previous findings that the line-of-sight to the Crab is underabundant
in oxygen, although more-so than recently measured. Using the abundances and 
cross sections from Wilms, Allen \& McCray (2000) we find [O/H] = ($3.33\pm
0.25$)$\times 10^{-4}$.
Analysis of the spectrum as a function of pulse phase measures the low-energy
X-ray spectral index even at pulse minimum~--- albeit with large statistical
uncertainty and we find marginal evidence for variations of the spectral index.
The data are also used to set a new ($3-\sigma$) upper limit to the temperature
of the neutron star of $\log T_{\infty} < 6.30$.
\end{abstract}


\section{Introduction} \label{s:intro}

The Crab Nebula and Pulsar constitute an intricate system, observed throughout
the electromagnetic spectrum.
Due to the complex X-ray structure of the inner nebula and pulsar, the
unprecedented angular resolution of the {\sl Chandra X-ray Observatory} has
proven invaluable in probing the nature of this region.
In Weisskopf et al.\ (2000), we presented a HETGS (High-Energy Transmission
Grating Spectrometer) zeroth-order image showing the complex morphology of the
inner nebula, revealing the previously undiscovered X-ray inner ring between
the pulsar and the X-ray torus.
In Tennant et al.\ (2001), our phase-resolved analysis of an LETGS (Low-Energy
Transmission Grating Spectrometer) zeroth-order image discovered significant
X-ray emission from the pulsar in its ``off'' phase.
Here we report our analysis of an LETGS dispersed image, to obtain
phase-resolved X-ray spectroscopy of the Crab Pulsar.

After briefly describing the observation and data reduction (\S \ref{s:obs}), we
discuss the analysis of the measured spectrum (\S \ref{s:res}).
In particular, we address photoelectric absorption and interstellar abundances
(\S \ref{s:ia}), impacts on the spectroscopy after allowing for scattering and
various abundances and cross-sections (\S \ref{s:fa}), comparison of our
results with certain other measurements (\S \ref{s:comp}), variation of the
nonthermal spectrum with pulse phase (\S \ref{s:var}), and constraints on the
temperature of the underlying neutron star (\S \ref{s:temp}).
Finally, we briefly summarize the results (\S \ref{s:sum}).

\section{Observation and Data Reduction} \label{s:obs}

On 2000 February 2, we obtained a nominally 50-ks observation (ObsID 759) of the
Crab Pulsar, with \cha 's Low-Energy Transmission Grating (LETG) and
High-Resolution Camera spectroscopy detector (HRC-S)~--- the LETGS. 
We examined the data after \cha\ X-ray Center (CXC) pipeline processing and
sorted events into an image binned in HRC pixels.
Using LEXTRCT (developed by one of us, AFT) we extracted the pulsar's dispersed
spectrum from the image.
The extraction uses a 29-pixel-wide (cross-dispersive) band centered on the
pulsar, in 2-pixel (dispersive) increments. 
For the LETG's 0.9912-$\mu$m grating period and 8.638-m Rowland-circle radius,
the LETGS dispersion is 1.148 \AA /mm.
Consequently, the spectral resolution of the binned data (two 6.4294-$\mu$m
HRC-S pixels) is 0.01476 \AA .
We combined positive and negative orders.

For estimating background, we similarly extracted reduced data from two
100-pixel-wide bands, starting 30 pixels to either side of the pulsar's
dispersed image (Fig.~\ref{f:regions}).
We restricted all spectral analysis to the (first-order) energy range 0.3-to-4.2
\kev.
The upper spectral limit avoids contamination from the zeroth-order nebular
image; the lower limit minimizes contamination from higher orders. 

\clearpage

\vspace{10pt}
\begin{center}
\includegraphics[angle=-90,width=\columnwidth]{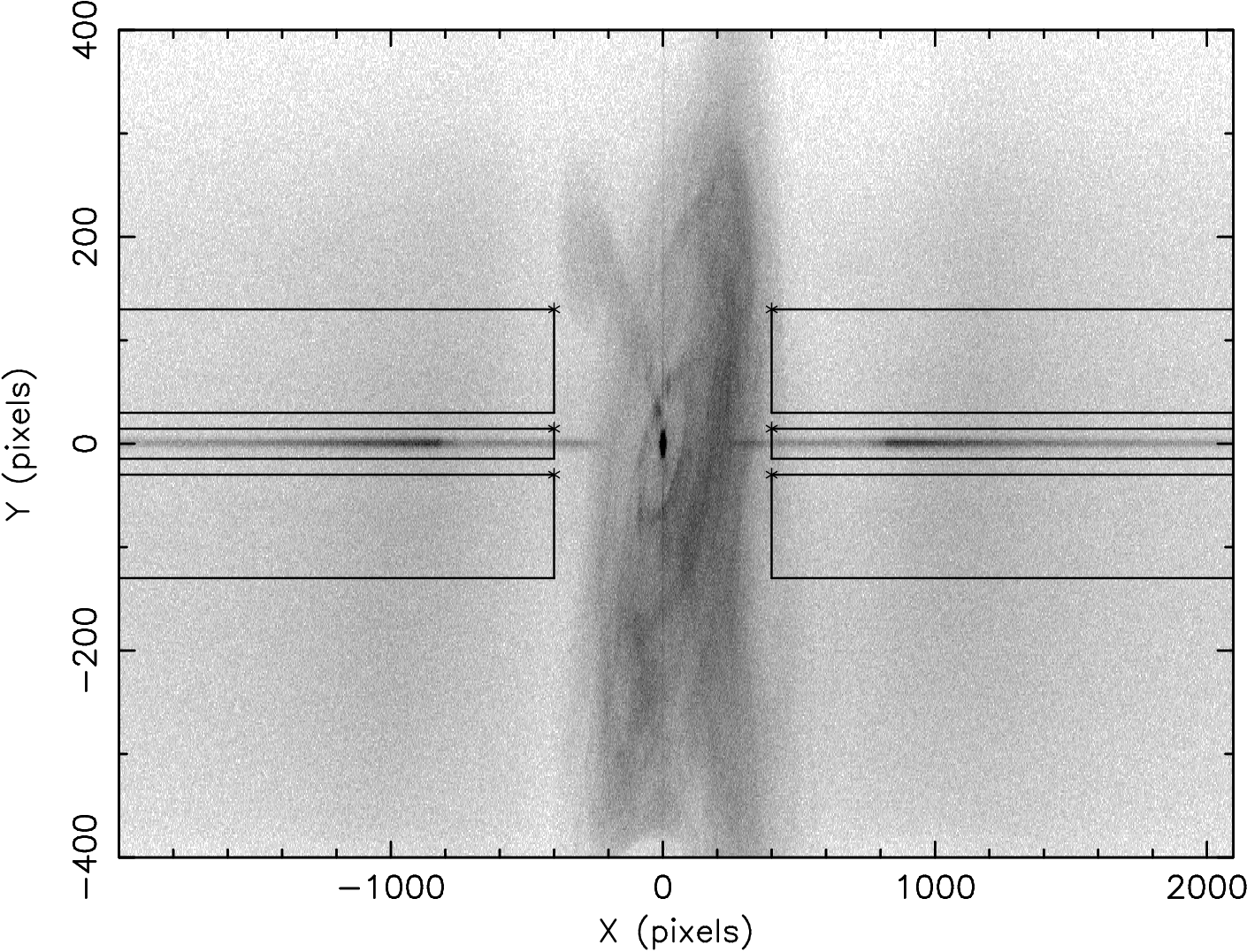}
\vspace{10pt}
\figcaption{LETGS image of the Crab showing the dispersed spectrum.
The rectangular boxes (discussed in the text) delineate regions selected for
analysis.
The image is stretched vertically for clarity.
\label{f:regions}
}
\end{center}

\clearpage

\section{Analysis and Results} \label{s:res}

We analyzed the data using the XSPEC (v.11.2) spectral-fitting package (Arnaud
1996).
To ensure applicability of the $\chi^{2}$ statistic, we merged spectral bins as
needed to obtain at least 100 counts per fitting bin (before background 
subtraction).
The merging results in no change in spectral resolution for the data above 0.67
keV and the merging of no more than three bins for the data above 0.5 keV.
We utilized an effective area (Fig.~\ref{f:ea}) that includes the LETG
energy-dependent efficiency to 10 spectral orders (Jeremy Drake, private
communication; see also http://asc.harvard.edu/cal/Letg/).

\clearpage

\vspace{10pt}
\begin{center}
\includegraphics[angle=-90,width=\columnwidth]{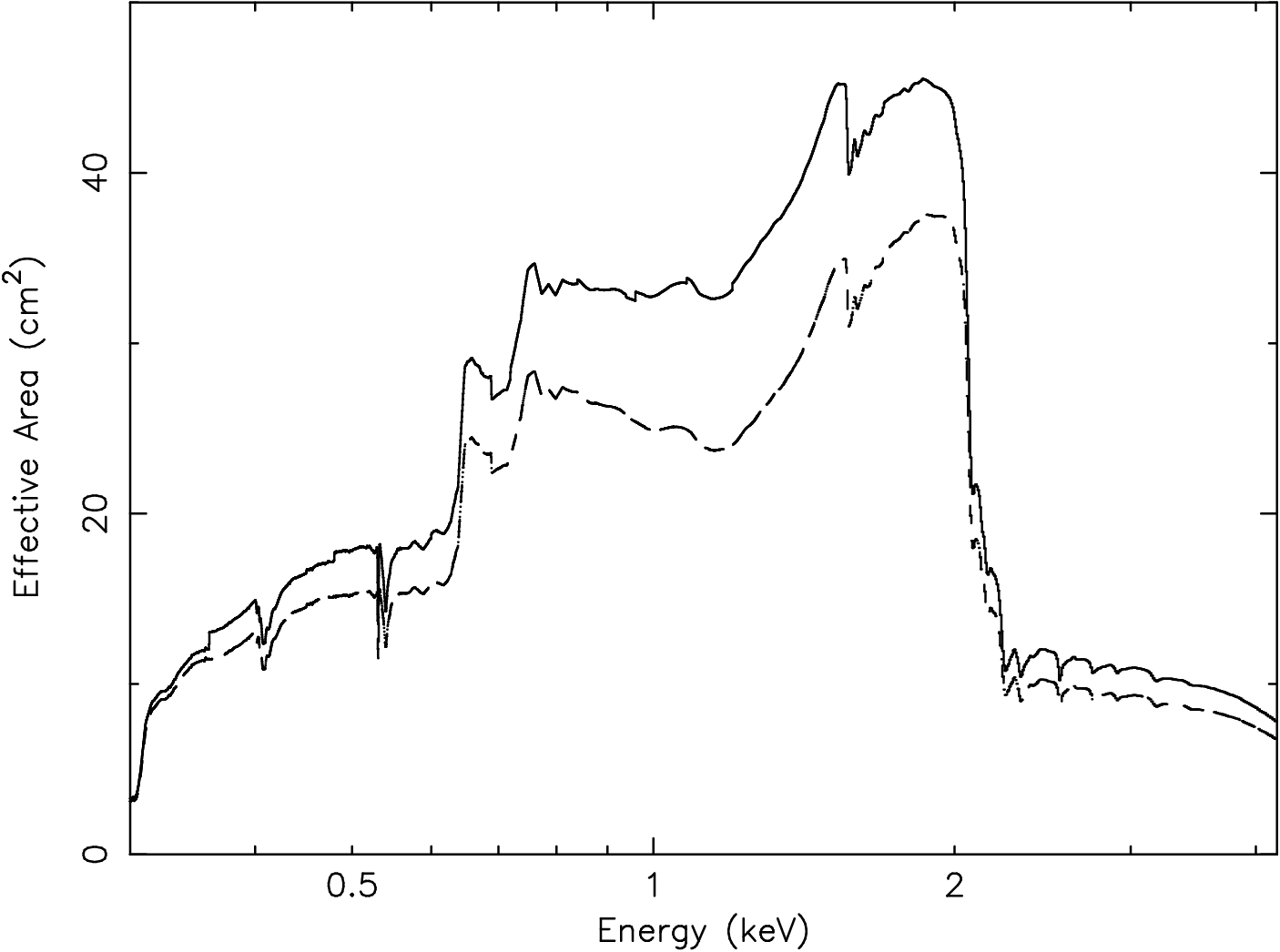}
\vspace{10pt}
\figcaption{LETGS effective area versus energy over the band of interest.
The solid line sums 10 (+ and -) spectral orders; the dashed line is the
first-order contribution.
 \label{f:ea}
}
\end{center}

\clearpage

\subsection{Broadband Spectrum~--- Initial Analysis} \label{s:ia}

\begin{figure}
\plotone{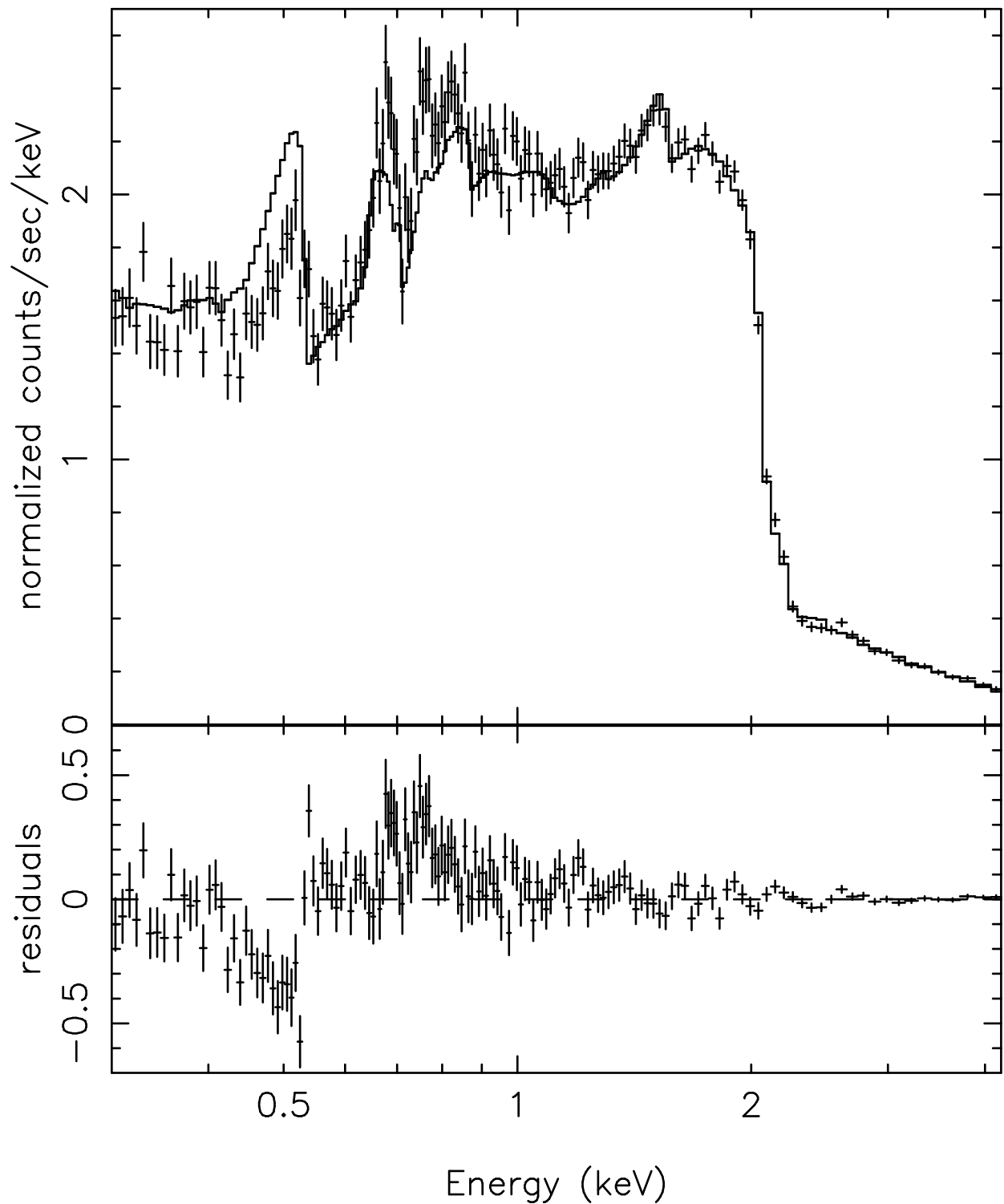}
\caption{\cha -LETGS spectrum of the Crab Pulsar compared to Model 1 (XSPEC {\tt
angr} abundances and {\tt bcmc} cross-sections).
Note the large residuals near the ${\rm O_{K}}$ edge (0.532 \kev) that result in
an unacceptable fit~--- $\chi^2$ = 1722 on 1553
degrees of freedom.
\label{f:model1}}
\end{figure}

We first used XSPEC to fit the spectral data, independent of pulse phase, in the
0.3--4.2-\kev\ band, using a power-law model with abundances {\tt angr} (from
Anders \& Grevesse 1989) and absorption cross-sections {\tt bcmc} (from
Baluncinska-Church \& McCammon 1992, with He cross-section from Yan,
Sagdeghpour, \& Dalgarno 1998).
Previous analyses of X-ray spectra~--- including the recent analysis of \xmm\
EPIC-MOS observations of the Crab
(Willingale et al.\ 2001)~--- use this combination (XSPEC {\tt angr} \& {\tt
bcmc}), which we denote as ``Model~1''.
The fit of the \cha\ LETGS data to Model~1 (Fig.~\ref{f:model1};
Table~\ref{t:sp-ave} entry 1) is statistically unacceptable,
yielding $\chi^2$ = 1722 on 1553 degrees of freedom.

\begin{figure}
\plotone{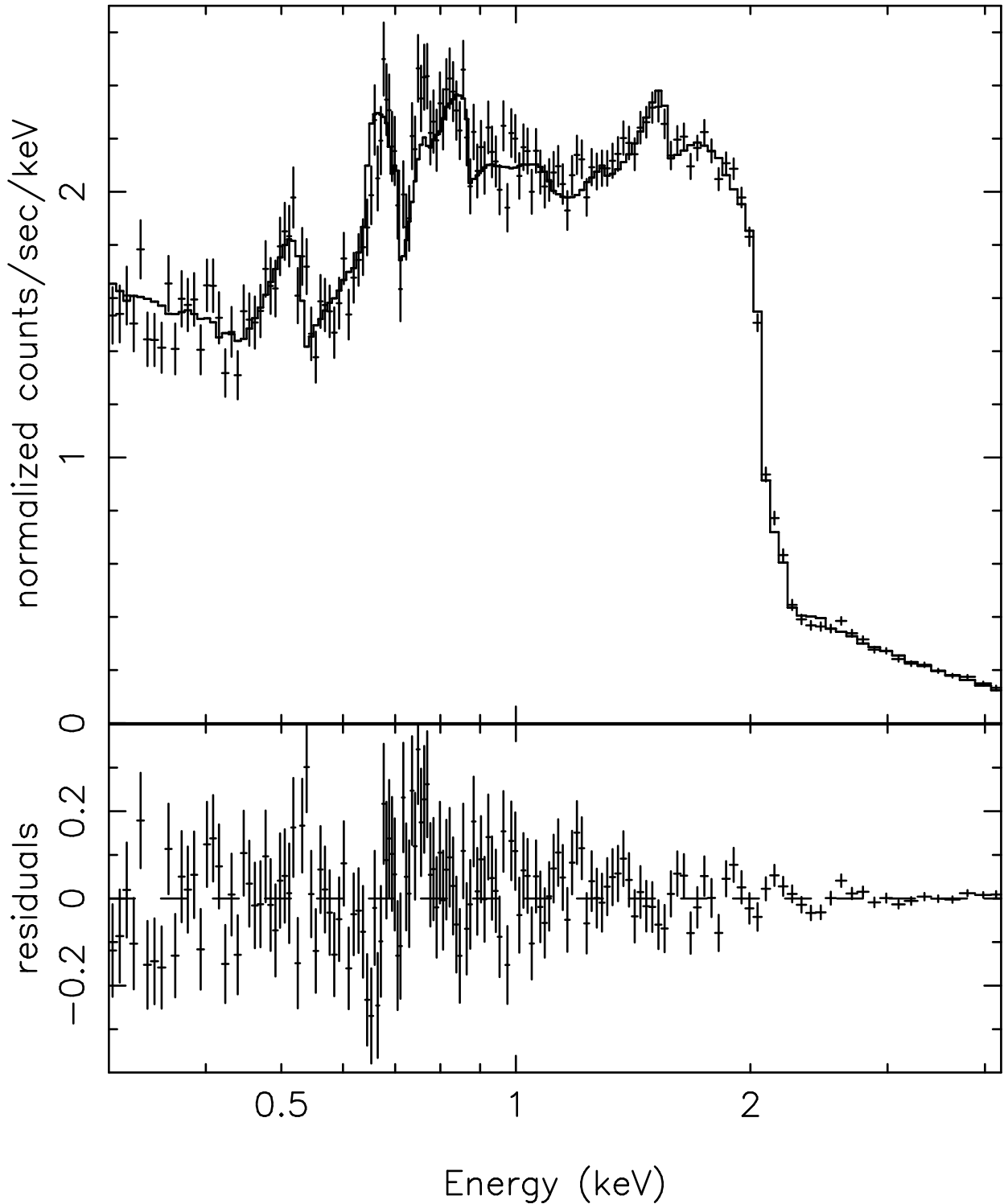}
\caption{\cha -LETGS spectrum of the Crab Pulsar compared to Model 2 (XSPEC {\tt
angr} abundances and {\tt bcmc} cross-sections
with thawed oxygen abundance). 
An acceptable fit~--- $\chi^2$ = 1546 on 1552 degrees of freedom~--- results if
oxygen is underabundant with respect to {\tt angr}
values.
 \label{f:model2}}
\end{figure}

Recognizing that the Model-1 residuals are largest near the ${\rm O_{K}}$ edge
(0.532~\kev), we thawed the relative abundance
of oxygen, holding fixed that of the remaining elements.
This ``Model~2'' (XSPEC {\tt angr} \& {\tt bcmc}, with thawed O abundance) leads
to a statistically acceptable fit
(Fig.~\ref{f:model2}; Table~\ref{t:sp-ave} entry 2)~--- $\chi^2$ = 1546 on 1552
degrees of freedom.
Table~\ref{t:sp-ave} lists the best-fit parameters and associated (1-$\sigma$)
statistical errors, determined from extrema on
single-parameter confidence contours.
Because statistical parameter estimation is valid only if the null hypothesis is
true, we omit ``best-fit'' parameters and errors
for poor fits.

The Model-2 best-fit power-law photon index for the pulsar~--- $\Gamma_{\rm P} =
1.596\pm0.020$~--- is less (i.e.,
harder) than for the nebula and similar to previous measurements ($\Gamma_{\rm
P} = 1.5 \pm 0.1$, Toor \& Seward 1974, 1977; 
$\Gamma_{\rm P} = 1.60\pm0.02$, Massaro et al.\ 2000) of the Crab Pulsar
spectrum.
Before comparing in more detail (\S \ref{s:comp}) our results with those of
other observations, we first examine
(\S\ \ref{s:fa}) the influence on spectral fits of using different cross-section
and absorption models
within XSPEC and of including effects due to scattering by interstellar grains. 

\clearpage

\clearpage

\clearpage

\begin{table}
\begin{center}
{\caption {XSPEC fits to \cha\ LETGS (phase-integrated) spectrum of the Crab
Pulsar.  \label{t:sp-ave}}}
\begin{tabular}{|cccccc|} 
\hline \hline
Abundances$^{a}$& Cross-sections$^{b}$ & $\chi^{2}$/$\nu$ & $\Gamma_{\rm P}$ &
$N_{\rm H}$   & [O/H] \\ 
               &                       &                  &                 
&\nsh           & $10^{-4}$    \\ \hline
 \multicolumn{6}{|c|}{Spectral fits without dust scattering} \\
{\tt angr}     & {\tt bcmc}            & 1772/1553$^{c}$  & $^{e}$           &
$^{e}$        & $8.51^{f}$   \\ 
{\tt angr}     & {\tt bcmc}            & 1546/1552$^{d}$  & $1.596\pm0.020$  &
$4.14\pm0.12$ & $3.45\pm0.26$\\ 
{\tt angr}     & {\tt obcm}            & 1724/1553        & $^{e}$           &
$^{e}$        & $8.51^{f}$   \\
{\tt angr}     & {\tt obcm}            & 1542/1552        & $1.600\pm0.020$  &
$3.77\pm0.11$ & $3.48\pm0.30$\\
{\tt angr}     & {\tt vern}            & 1559/1552        & $1.611\pm0.020$  &
$4.33\pm0.12$ & $3.93\pm0.25$\\
{\tt wilm}     & {\tt vern}            & 1561/1553        & $1.549\pm0.019$  &
$4.39\pm0.08$ & $4.90^{f}$   \\  
{\tt wilm}     & {\tt vern}            & 1540/1552        & $1.541\pm0.019$  &
$4.84\pm0.15$ & $3.47\pm0.24$\\
{\tt wilm}     & {\tt tbvarabs}        & 1551/1553        & $1.543\pm0.019$  &
$4.24\pm0.08$ & $4.90^{f}$   \\  
{\tt wilm}     & {\tt tbvarabs}        & 1540/1552        & $1.536\pm0.019$  &
$4.55\pm0.15$ & $3.76\pm0.35$\\ \hline
 \multicolumn{6}{|c|}{Spectral fits with dust scattering --- $\tau_{\rm
scat}=0.15$ at 1 \kev} \\
{\tt angr}     & {\tt bcmc}            & 1766/1553        & $^{e}$           &
$^{e}$        & $8.51^{f}$   \\ 
{\tt angr}     & {\tt bcmc}            & 1555/1552        & $1.643\pm0.020$  &
$3.77\pm0.12$ & $3.18\pm0.30$\\ 
{\tt angr}     & {\tt obcm}            & 1722/1553        & $^{e}$           &
$^{e}$        & $8.51^{f}$   \\
{\tt angr}     & {\tt obcm}            & 1551/1552        & $1.647\pm0.020$  &
$3.43\pm0.11$ & $3.20\pm0.34$\\
{\tt angr}     & {\tt vern}            & 1572/1552        & $1.655\pm0.020$  &
$3.96\pm0.12$ & $3.56\pm0.30$\\
{\tt wilm}     & {\tt vern}            & 1571/1553        & $1.601\pm0.019$  &
$3.93\pm0.08$ & $4.90^{f}$   \\  
{\tt wilm}     & {\tt vern}            & 1540/1552        & $1.592\pm0.019$  &
$4.46\pm0.15$ & $3.08\pm0.28$\\ 
{\tt wilm}     & {\tt tbvarabs}        & 1558/1553        & $1.597\pm0.019$  &
$3.80\pm0.08$ & $4.90^{f}$   \\  
{\tt wilm}     & {\tt tbvarabs}        & 1539/1552$^{g}$  & $1.587\pm0.019$  &
$4.20\pm0.14$ & $3.33\pm0.25$\\ \hline
\end{tabular}
 \tablecomments{\\
$^{a}$ Abundance models in XSPEC: {\tt angr}, Anders \& Grevesse (1989); {\tt
wilm}, Wilms, Allen, \& McCray
(2000)\\
$^{b}$ Cross-section models in XSPEC: {\tt bcmc}, Balucinska-Church \& McCammon
(1992) with He cross-section from
Yan, Sadeghpour, \& Dalgarno (1998); {\tt obcm}, Balucinska-Church \& McCammon
(1992); {\tt vern}, Verner et al.\
(1996); {\tt tbvarabs}, Wilms et al.\ (2000, references therein) allowing for
absorption by interstellar grains \\
$^{c}$ Model 1 (see text and Fig.~\ref{f:model1})\\
$^{d}$ Model 2 (see text and Fig.~\ref{f:model2})\\
$^{e}$ ``Best-fit'' parameters and errors omitted for statistically poor fits\\
$^{f}$ Default relative abundance of oxygen for the given abundance model\\
$^{g}$ Model 3 (see text and Fig.~\ref{f:model3}), adopted for subsequent
analyses\\
}
\end{center}
\end{table}

\clearpage

\subsection{Broadband Spectrum~--- Further Analysis} \label{s:fa}

Beyond the initial analysis (\S \ref{s:ia}), we investigated effects on model
fits and parameters using various combinations of cross-section and abundance
models available within XSPEC.
Besides the abundance model {\tt angr} (\S \ref{s:ia}), we considered {\tt wilm}
(Wilms, Allen, \& McCray 2000).
Besides the cross-section model {\tt bcmc} (\S \ref{s:ia}), we considered {\tt
obcm} (Baluncinska-Church \& McCammon 1992,
with old He cross-section), {\tt vern} (Verner et al.\ 1996), and {\tt tbvarabs}
(Wilms et al.\ 2000).
Wilms et al.\ (2000) employ updated abundances (Snow \& Witt 1996; Cardelli et
al.\ 1996; Meyer et al.\ 1997, 1998) and
recent theoretical results for elemental absorption cross-sections (Band et al.\
1990; Yan et al.\ 1998; Verner et al.\
1993; Verner \& Yakovlev 1995) and for molecular hydrogen (Yan et al.\ 1998).
In addition, their cross-section models (XSPEC's {\tt tbabs} and {\tt tbvarabs})
include effects on interstellar
absorption due to condensation into grains, which becomes important for grains
sufficiently large to be opaque at a given energy. 

Owing to the small effective aperture of our observation, we deemed it prudent
also to consider effects of (diffractive) scattering
by grains upon interstellar extinction.
Such scattering produces a wavelength-dependent X-ray scattering halo (Overbeck
1965; Mauche \& Gorenstein 1986), related to the
diffractive (half-power) angle $\vartheta_{\rm HP} \approx 70 \arcsec
/[(a/{\mu\rm m})(E/{\rm ke\!V})]$.
Consequently, for energies of interest here, all but the largest grains scatter
outside the source extraction aperture, thus
contributing to extinction.
(See Takei et al.\ 2002 for another example of this effect.)
Following Mauche \& Gorenstein (1986), we calculate this scattering in the
Rayleigh--Gans approximation (van der Hulst 1957; Overbeck 1965; Hayakawa
1970), valid when the phase shift through a grain diameter is small.
This condition limits applicability to grains with radii
\begin{equation} \label{e:rg_max-a}
a << (0.5\, \mu{\rm m}) \left( {E}\over{\rm{ke\!V}} \right) \left({{\rho_{\rm 
grain}}\over{{\rm g\, cm}^{-3}}}\right)^{-1} \left( {2 Z F(E)}\over{M}
\right)^{-1} ,
\end{equation}
with $\rho_{\rm grain}$ the mass density of a dust grain and $Z$ and $M$ the
summed atomic number and weight,
respectively, of a grain-material molecule.  
The near-unity function $F(E) \equiv | \sum_{i} N_{i} f_{i}(E) | / (\sum_{i}
N_{i} Z_{i})$,
with $f_{i}(E)$ the (complex) atomic scattering factor of atomic species $i$,
$N_{i}$ the number of species-$i$ atoms
per molecule, and $Z_{i}$ the species-$i$ atomic number.  

In the Raleigh-Gans approximation, the scattering depth is (Mauche \& Gorenstein
1986)
\begin{equation} \label{e:tau-scat}
\tau_{\rm scat}(E) \equiv \tau_{\rm s1} \left( {F(E)}\over{F(1\, {\rm
ke\!V)}}\right) ^{\! 2} 
\left( {E}\over {\rm ke\!V}\right) ^{\! -2}\ ,
\end{equation}
with $\tau_{\rm s1}$ the dust scattering depth at 1~\kev. 
For computations, we calculate $F(E)$ for the dust composition implied by the
abundances and dust depletion factors
given in Table 2 of Wilms et al.\ (2000), using atomic scattering factors from
Henke et al.\ (1993).
For the relevant case, which is extraction-aperture limited, a suitable fitting
function, $N_{E}(E)$, for an power-law photon index spectrum is
\begin{equation} \label{e:spec}
N_{E}(E) = {\rm constant}\times E^{-\alpha} \exp(-\tau_{\rm abs}(E))
\exp(-\tau_{\rm scat}(E))\ ,
\end{equation}
with $\tau_{\rm abs}(E)$ the photoelectric absorption depth and $\tau_{\rm
scat}(E)$ the scattering depth (Eq.~\ref{e:tau-scat}).  

\begin{figure}
\plotone{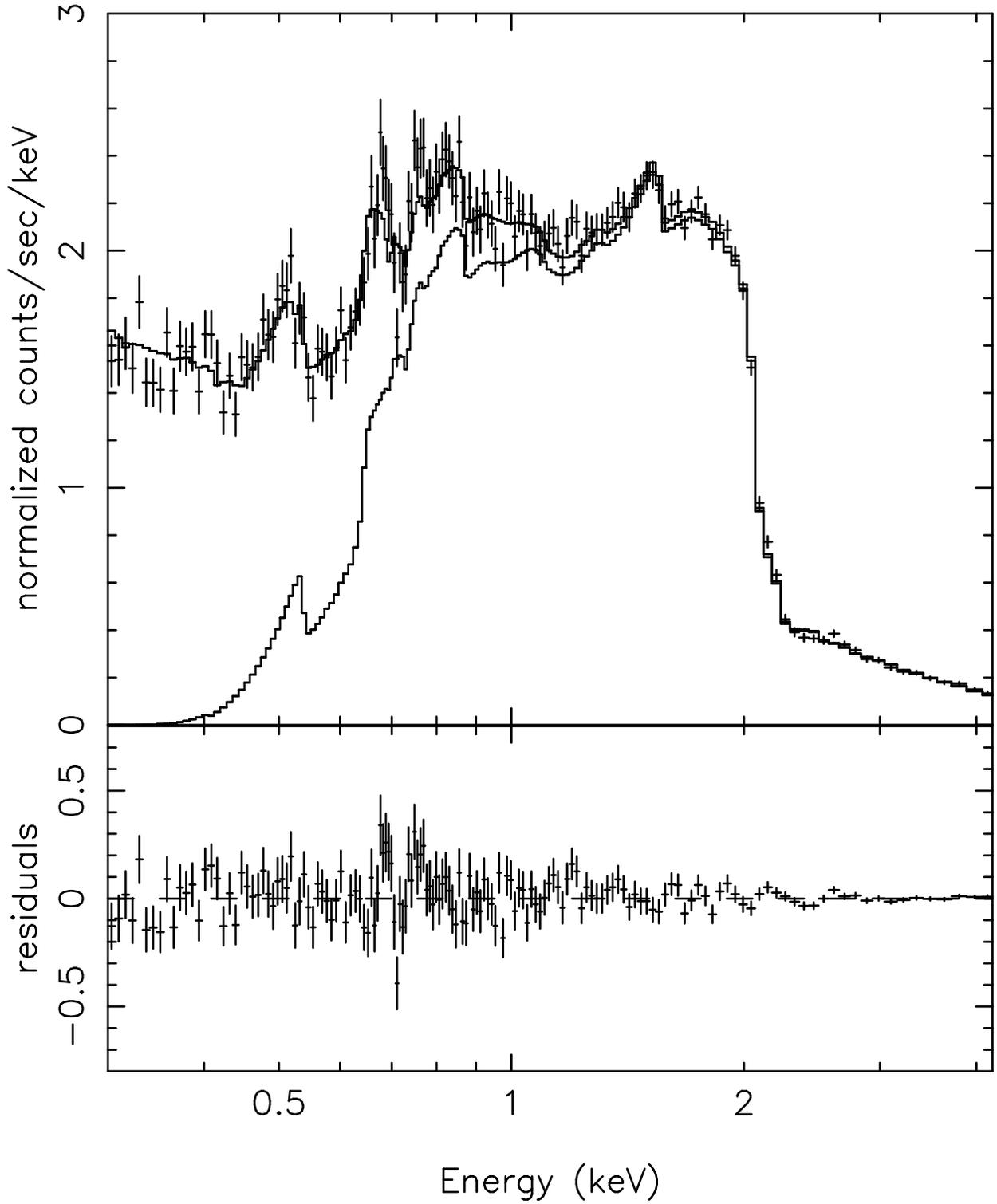}
\caption{\cha -LETGS spectrum of the Crab Pulsar compared to Model 3 (XSPEC {\tt
wilm} \& {\tt tbvarabs} with thawed oxygen abundance and with $\tau_{\rm s1} =
0.15$, last entry in Table~\ref{t:sp-ave}). 
This represents the preferred model, which serves as the basis for subsequent
analyses. The lower solid line shows the model including only the first order
response.
\label{f:model3}}
\end{figure}

For purposes of this study, we set $\tau_{\rm s1} = 0.15$ and assume validity of
the Rayleigh-Gans approximation, consistent with
measurements of the Crab scattering halo (Mauche \& Gorenstein 1989; Predehl \&
Schmitt 1995).
We did not investigate more sophisticated grain-scattering models, because using
the simple scattering model (Eq.~\ref{e:tau-scat})
or ignoring scattering completely didn't affect the quality of the fits, most
of which were statistically acceptable.
Table~\ref{t:sp-ave} lists best-fit parameters and statistical errors (where the
fit was statistically acceptable) for models
ignoring and those including extinction due to scattering by interstellar
grains.
The first two table entries are Model 1 (Fig.~\ref{f:model1}) and Model 2
(Fig.~\ref{f:model2}) of the initial analysis (\S \ref{s:ia}).
The last entry is Model 3 (Fig.~\ref{f:model3}) which we consider to be
astrophysically most accurate and thus adopt in subsequent analyses (\S
\ref{s:var} and \S \ref{s:temp}). 

\clearpage

\clearpage
  
Inspection of Table~\ref{t:sp-ave} allows an assessment of the influence of the
model upon the determination of the relevant
parameters. 
In particular, we note the following: 
\begin{enumerate}
  \item Independent of the cross-section model, it is impossible to fit the data
using the abundances of Anders \& Grevesse
(1989). 
  \item Models invoking the Anders \& Grevesse (1989) abundances will fit the
data if the oxygen is underabundant with a fractional abundance of $0.41 \pm 
0.07$. 
  \item If abundances recommended by Wilms et al.\ (2000) are used, the fits
become acceptable independent of
the various cross-sections used. 
  \item In all cases with statistically acceptable fits, the best-fit relative
oxygen abundance is basically the same. 
  \item Using the Wilms et al.\ (2000) abundances with oxygen somewhat
underabundant consistently gives a better fit than if the abundance is fixed at
the nominal value.
  \item Including extinction by dust scattering does not appreciably alter the
quality of the fit; thus, while such extinction is expected, these spectral data
do not require it.  
  \item Including dust scattering consistently modifies the best-fit parameters
in an understandable way~--- the resulting photon
index $\Gamma_{\rm P}$ is slightly flatter and the hydrogen column $N_{\rm H}$
and relative oxygen abundance [O/H] are slightly lower.
  \item The range of best-fit values for $N_{\rm H}$ substantially exceeds the 
statistical error of that parameter, demonstrating the importance of systematic
uncertainties produced by using different cross-sections and/or abundances.
\end{enumerate}

\clearpage
  
\subsection{Broadband Spectrum~--- Comparison with Certain Previous
Observations}
\label{s:comp}

Here we compare our measurements of the broadband spectrum with certain previous
measurements~--- those taken with the \ein\ Focal-Plane Crystal Spectrometer
(FPCS) by Schattenburg \& Canizares (1986), with the \sax\ Low-Energy
Concentrator Spectrometer (LECS) by Massaro et al.\ (2000), and with the \xmm\
European Photon Imaging Camera (EPIC) by Willingale et al.\ (2001). 

\subsubsection{The \ein\ measurements} \label{s:einstein}

Using the \ein\ FPCS, Schattenburg \& Canizares (1986) measured the spectrum of
the Crab Nebula and its the pulsar,
finding $N_{\rm H} = (3.45\pm0.42)\! \times\! 10^{21}\ {\rm cm}^{-2}$ and
$N_{\rm O} = (2.78\pm0.42)\! \times\! 10^{18}\ {\rm cm}^{-2}$, giving a
relative oxygen abundance $[O/H]=(8.1\pm1.6)\! \times\! 10^{-4}$. 
Because the measurements included all flux in a $6\arcmin$ field of view, they
did did not resolve the pulsar from the nebula. 
Their analysis did allow for the impact of flux scattered outside of the
spectrometer's 6\arcmin-diameter aperture upon the overall
normalization. 
The model fitting of the \ein\ data froze the nebula's power-law index at
$\Gamma_{\rm N} = 2.1$ and used cross-sections and abundances from
Morrison \& McCammon (1983), but did not report the goodness of the fit.

If the quality of the fit were poor, statistical uncertainties in $N_{H}$ and
$N_{O}$ may be underestimated, although the authors noted that they
deliberately inflated the statistical errors in the data above 0.7~\kev\ in
their spectral fitting, perhaps to avoid this problem. 
In what follows, we shall first assume that the uncertainties are not
underestimated. 
Before comparing results we note that their freezing the spectral index lowers
the uncertainty relative to a fit with the spectral index left as a free
parameter, but the differences may not be very large and their choice of
spectral index (2.1) appears to be a very accurate value for the
nebula-averaged index at low X-ray energies (see the discussion in \S
\ref{s:xmm}). 
Similarly, we find for the \cha\ data, e.g., that the uncertainty in $N_{\rm H}$
and $N_{\rm O}$ change by less than 10\% depending on whether $\Gamma$ is left
as a free parameter or fixed at the best-fit value. 
Thus, if we compare the \ein-derived values for the hydrogen column and the
relative oxygen abundance with our measurements, we see that the values for 
$N_{H}$ are similar within the errors, essentially independent of which
cross-sections and abundances we used, whereas there appears to be a
statistically significant discrepancy in the relative oxygen abundance. We
shall see shortly (\S~\ref{s:bepposax} \& \S~\ref{s:xmm}) that this 
discrepancy is with all the more recent measurements. 
Moreover, Willingale et al.\ (2001) find that the column is rather uniform
across the nebula (see \S \ref{s:xmm}), thus eliminating variations in the
column as a possible reason for the discrepancy. 
  
In order to try to understand the source of this discrepancy, we also compared
the measured optical depths at the oxygen edge, which are directly related to
the data independent of any models for abundance and cross-section.   

Following Schattenburg \& Canizares (1986), we omitted 2 of the 6 FPCS data
points that lie closest to the oxygen edge as outliers, which they note are
perhaps biased by neutral-oxygen ${\rm K}_\alpha$ absorption and by the {\rm K}
edge of singly ionized oxygen. 
Scaling their quoted uncertainty in the fractional abundance of
oxygen, we find $\tau_{\rm O} = 1.6\pm0.3$.
If we include all 6 FPCS data points in the analysis, we find $\tau_{\rm O} =
0.8$, with a very poor fit ($\chi^{2} = 15$ for 3 degrees of freedom).
This leads us to conclude that had they included these points the value would
have been smaller and the uncertainty in the optical depth larger. 
By way of comparison, we find for the \cha -LETGS data that  $\tau_{\rm O} =
0.64\pm0.06$ in the direction of the pulsar.
We also see no reason as to why the systematic errors that Schattenburg \&
Canizares (1986) applied to the data above 0.7 keV do not also apply below 0.7
keV. 
This would surely increase the uncertainty.
As we have shown the FPCS result is sensitive to precisely which data is
included in the fit, and, we suspect, to the precise modelling of the details
of the spectral structure. 
We can only conclude that either the \ein-FPCS uncertainties in the measured
oxygen column (relative oxygen abundance) are underestimated or some systematic
effect in the response has escaped detection. 

\subsubsection{The \sax\ measurement} \label{s:bepposax}

Massaro et al.\ (2000) used various \sax\ instruments to measure the spectrum of
the nebula and pulsar, which are spatially unresolved even using the \sax\
X-ray concentrators.
Of relevance to our measurements, the Low-Energy Concentrator Spectrometer
(LECS) obtained a phase-resolved 0.1--4.0-\kev\ spectrum of the nebula and
pulsar. 
By analyzing the spectrum from the ``low-pulse'' phases, they minimize the
pulsar contribution relative to the nebula. 

T. Mineo kindly re-analyzed the \sax\ LECS data, finding that underabundant
oxygen substantially improved the fit ($\Delta \chi ^2 = 48$ for 1 additional
parameter) using the XSPEC {\tt angr} abundances and {\tt bcmc} cross-sections
(cf.\ the second model of Table~\ref{t:sp-ave}).
This new fit gives $\Gamma_{\rm N} = 2.06\pm0.08$, $N_{\rm H} = (3.54\pm0.07)
\times 10^{21}\; {\rm cm}^{-2}$, and $[O/H] = (6.25\pm0.33) \times
10^{-4}$; moreover, the fit was now marginally statistically acceptable with
$\chi^2 = 428$ on 359 degrees of freedom. 
Thus, re-analysis of the \sax\ data also indicates that oxygen in the
Crab line of sight to be underabundant with respect to Anders \&Grevesse (1989)
abundances, however, not as much as our results. 

\subsubsection{The \xmm\ measurement} \label{s:xmm}

Our final comparison is with the \xmm\ observations reported by Willingale et
al.\ (2001).
Their analysis used 0.4--3.0-\kev\ MOS-1 and -2 data, including data from the
trailed (``out of time'') image, to  fit five different data sets. 
The selected data sets isolated various Crab spatial structures~--- including
the jet, the nebula as a whole, and the pulsar. 
The joint spectral fits to these data sets employed different power-law indices
for each region but the same column density, with cross-sections from
Balucinska-Church \& McCammom (1992) plus newer He cross-section (Yan et al.\
1998) and with abundances from Anders \& Grevesse (1989).

Willingale et al.\ (2001) found statistically unacceptable fits before thawing
both oxygen and iron abundances.
Under the assumption that O and Fe have the same fractional abundance relative
to solar, they determined the fractional abundance
to be $0.63\pm0.01$ times solar (Anders \& Grevesse 1989).
These authors noted that the  conclusion that oxygen may be underabundant is in
agreement with earlier inferences from {\sl Copernicus} data (e.g. Keenan,
Hibbert, and Dufton 1985 and
references therein).
The best-fit parameters were $\Gamma_{\rm N} =
2.108\pm0.006$, $\Gamma_{\rm P} = 1.63\pm0.09$, $N_{\rm H} = (3.45\pm0.02)\times 
10^{21}\; {\rm cm}^{-2}$, and $[O/H] = (5.36\pm0.10) \times 10^{-4}$, with 
a  $\chi^{2} = 4007 $ on $\nu = 3853$ degrees of freedom.
In addition, by sampling $20\arcsec\! \times\!  20\arcsec$ regions, Willingale
et al.\ determined that column-density variations across the nebula are small.

We examined our data to determine whether we could confirm an underabundance 
of iron.
Using our preferred model  (Model 3, Table~\ref{t:sp-ave}), we thawed 5 XSPEC
parameters~--- the power-law index,  its normalization, the hydrogen column
density, the fractional abundance of O,  and individually the fractional
abundance of He, C, N, Na, Mg, Al, Si, Cl, Ar, Ca, Cr, Fe, and Co.  
Here the fractional abundance is the ratio of the abundance of a given element
to its standard (default) abundance.
For nearly all these  elements, the \cha -LETGS spectroscopy was insensitive up
to a fractional 
abundance from Wilms et al.(2000) of at least 2.  
The exceptions with regards to the sensitivity were helium and iron with
fractional abundances of $1.07\pm0.20$ and $0.77\pm0.18$, respectively. 
These to be compared with an oxygen fractional abundance of $0.68\pm0.05$
already discussed.
Therefore, in all cases, except for the oxygen, our results were consistent with
normal abundances relative to the reference. 
The uncertainty and the trend in the relative iron abundance is consistent with
it being either normal (with respect to Wilms et al.\ 2000) or low, and tied to
that of oxygen. 
We therefore conclude that our results are qualitatively consistent with those
of Willingale et al.\ (2001), although we disagree as to the degree of the
oxygen underabundance.  
There are no easy ways to account for the discrepancy.
Both analyses agree within statistical errors on the hydrogen column, when the
same models for cross sections and abundances are used.
Willingale et al.\ (2001) did not account for scattering by the ISM which tends
to lower the estimated relative oxygen abundance (see Table~\ref{t:sp-ave}) but
it is difficult to see that this would be a very large factor, especially
since the \xmm\ beam is so much larger than that of \cha. 
A significant difference between the two experiments is the spectral resolution,
in that Willingale et al.\ (2001) used CCDs. 
It is possible that the better \cha-LETGS spectral resolution, especially with
regards to details of the oxygen edge, might account for the discrepancy. 
It is of course possible that the \cha\ response is in error, but the good fit
of the pulsar spectrum to the powerlaw model seems to indicate that the model
is correct.
We note that the LETGS response was {\em not} calibrated using the Crab. 
Future experiments that deal with the accuracy of the calibration of the
instruments will be necessary to clarify the details. 

\clearpage\subsection{Spectral Variation with Pulse Phase} \label{s:var}

In principle, repeating the previous analysis for the time-tagged data
(appropriately arranged by pulse phase) provides a phase-resolved measurement
of the pulsar's spectral parameters. 
However, a HRC-S timing error assigns to each event the time of the previous
event, thus complicating the analysis for this bright source because telemetry
saturation omits events from the telemetry stream.
In Tennant et al.\ (2001), we discussed this problem and a method for
maintaining timing accuracy~--- albeit at significantly
reduced (here, by a factor of 15) data-collection efficiency.
This method filters the data accepting only telemetered events separated by no
more than 2 ms, guaranteeing a timing accuracy never
worse than 2 ms and typically much better.
Jodrell Bank (Lyne, Pritchard, \& Smith 1993) routinely observes the Crab Pulsar
(Wong, Backer,\& Lyne 2001); providing a period
ephemeris ({\tt http://www.jb.man.ac.uk/$\sim$pulsar/crab.html}). 
Roberts \& Kramer (2000, private communication) kindly prepared an ephemeris
matched to our observation times.
Applying the 2-ms filter and folding the data according to the radio ephemeris
we found that the X-ray flux peaks at phase 0.984 as discussed in Tennant et
al.\ (2001).

In performing the phase-resolved spectral analysis, we used interstellar
absorption and dust scattering parameters of our preferred XSPEC model (Model 3,
last entry of Table~\ref{t:sp-ave}and Figure~\ref{f:model3}).
 Table~\ref{t:phase} and Figure~\ref{f:lc_gamma} summarize the results for the
pulsar's phase-resolved photon index $\Gamma_{\rm P}$. 
Note that for each phase range, the power-law fit is statistically acceptable
without requiring an additional spectral component. 

\clearpage

\begin{table*}
\begin{center}
\caption {Spectral Parameters versus Pulse Phase. 
Uncertainties for the photon index $\Gamma_{\rm P}$ are 1-$\sigma$ statistical
errors. \label{t:phase}}
\small{
\begin{tabular}{|ccc|} 
\hline \hline
Phase range    & $\Gamma_{\rm P}$       & $\chi^{2}$/$\nu$   \\
\hline
   0.01--0.06  & $1.71\pm0.07         $ & 32.9/37 \\ 
   0.06--0.10  & $1.53\pm0.20         $ & 20.7/15 \\ 
   0.10--0.20  & $1.53\pm0.15         $ & 35.1/35 \\ 
   0.20--0.30  & $1.50\pm0.09         $ & 61.3/46 \\ 
   0.30--0.35  & $1.43\pm0.07         $ & 30.7/35 \\ 
   0.35--0.42  & $1.54\pm0.04         $ & 74.1/82 \\  
   0.42--0.45  & $1.72\pm0.11         $ & 16.6/18 \\ 
   0.45--0.55  & $1.78\pm0.16         $ & 39.0/34 \\ 
   0.55--0.70  & $1.4^{+1.0}_{-1.3}   $ & 36.2/41 \\ 
   0.70--0.82  & $2.9\pm1.0           $ & 34.9/34 \\ 
   0.82--0.90  & $1.47\pm0.63         $ & 35.6/24 \\ 
   0.90--0.95  & $1.48\pm0.08         $ & 29.9/31 \\
   0.95--0.01  & $1.61\pm0.03         $ & 132/105 \\ \hline
\end{tabular}
} 
\end{center}
\end{table*}

\clearpage

Based upon a $\chi^2$ analysis of the distribution of best-fit photon indices
(Table~\ref{t:phase}), we reject (albeit at only  85\% confidence)
the hypothesis that the spectral index is constant with phase. (The
error-weighted average
of the spectral indices was 1.58 and the value of $\chi^2$ was 16.8 on 12
degrees of freedom.)
The apparent variation of spectral index between pulse phases -0.1 and 0.5 is
qualitatively similar in \cha, \sax\
(Massaro et al.\ 2000), and \xte\ (Pravdo, Angelini, \& Harding 1997)
measurements, with the index increasing (becoming softer)
through the two pulse maxima and decreasing (becoming harder) in the bridge
between the pulses.
However, only \cha\ provides the angular resolution needed to isolate the pulsar
from the nebula in order to measure the
spectral index for the pulse-phase range 0.5--0.9. 
Our analysis is consistent with an apparent continuation of the increase
(softening) of the spectral index until just before the
onset of the primary pulse; however, the data do not require this. 
Due to the HRC time-tag problem, the efficiency of the \cha\ observation was low
(5.5\%); hence, the spectral-index
uncertainty near pulse minimum is large.
We have proposed a much more efficient approach for collecting \cha-LETGS data
from the Crab Pulsar, which would
significantly reduce statistical uncertainties without expending inordinate
observing time.
These future observations would directly challenge theoretical models for pulsar
emission (e.g., Zhang \& Chen 2002).

\clearpage

\vspace{10pt}
\begin{center}
\includegraphics[angle=-90,width=\columnwidth]{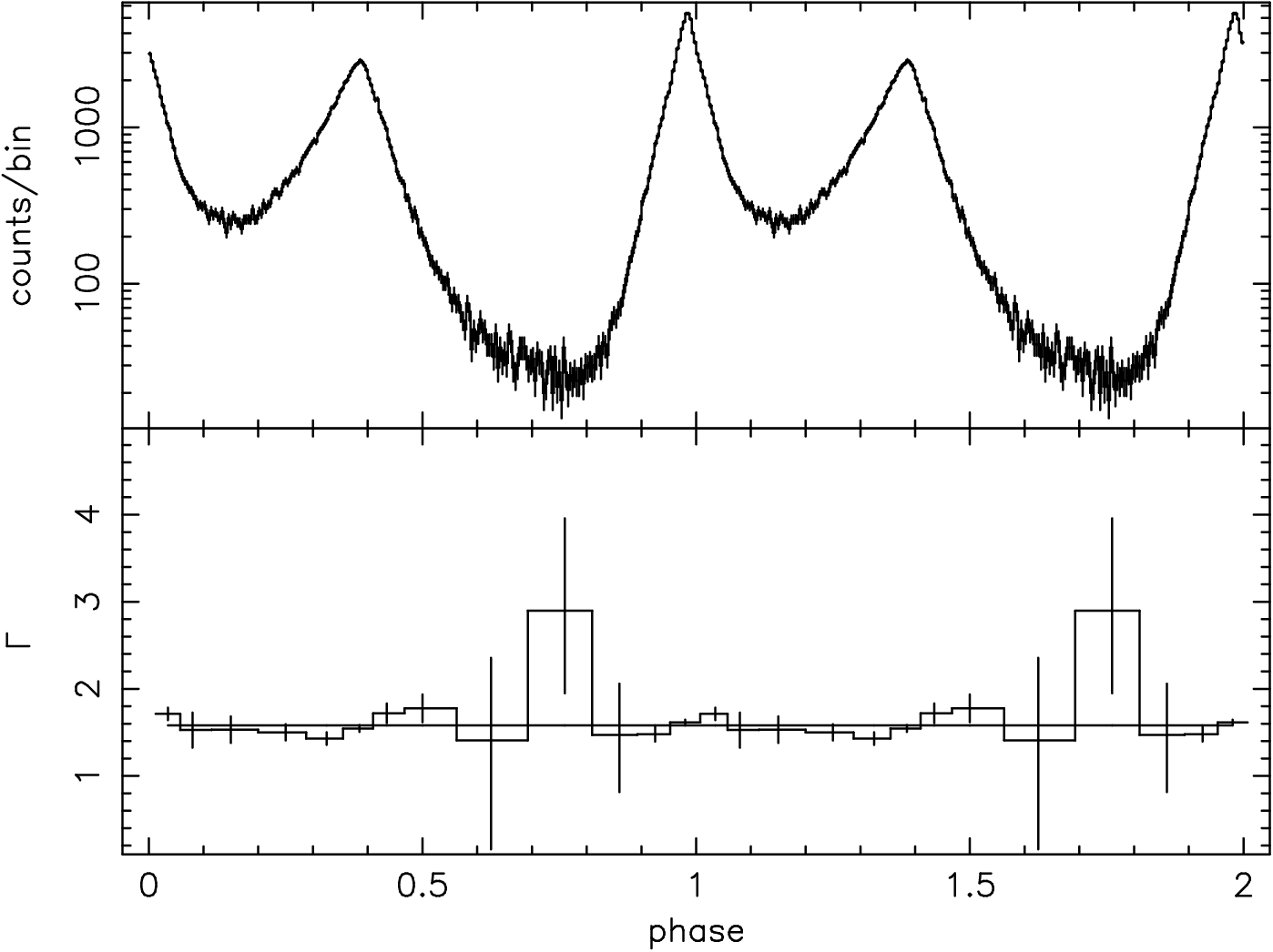}
\vspace{10pt}
\figcaption{Crab Pulsar light curve and photon index as a function of pulse
phase.
Note that the plots span two (identical) pulse cycles.
 \label{f:lc_gamma}}
\end{center}

\clearpage

\subsection{Temperature of the Neutron Star} \label{s:temp}

Here we investigate adding a blackbody component to the spectral fits in order
to set an upper limit to the temperature of the underlying neutron star. 
Ideally, data taken at pulse minimum would provide the most stringent upper
limit.
In practice, however, the phase-averaged data (which do not require
time-interval filtering) yield the best limit due to their much better photon
statistics.
Figure~\ref{f:bbnorm_kT} shows a portion of the confidence contours of the
blackbody normalization $(\theta_{\infty}^{2})$ versus blackbody temperature
($kT_{\infty}$) using the preferred model ({\tt wilm} and {\tt tbvrabs}) with
$\tau_{\rm s1} = 0.15$, but $\Gamma_{\rm P}$, $N_{\rm H}$, and [O/H] thawed. 
Here $\theta_{\infty}$ is the angular size determined by a distant observer, in
XSPEC units~--- $\theta_{\infty} = (R_{\infty}/D_{10})$, with $R_{\infty}$ the
stellar radius in km units and $D_{10}$ the source distance in 10-kpc units.

The best fit (not shown in the figure) gave $kT = 187$ \kev\ and
$\theta_{\infty}^{2}=0.0018, [{\rm km}/(10\, {\rm kpc})]^2 = [8.5\, {\rm
m}/(2\, {\rm kpc})]^2$, with $\chi^{2} = 1537$ on 1550 degrees of freedom. 
Clearly, the best-fit parameters~--- corresponding to a radius of less than 10
meters~--- cannot represent thermal emission from the entire surface of a
neutron star, although they might indicate a very small, very hot spot on the
surface.
It is more reasonable, however, that the best-fit high temperature and low
normalization are indicative of the absence of any thermal component.
Figure~\ref{f:bbnorm_kT} shows the upper portions of the ``banana'' plot for
contours consistent with blackbody emission from the entire neutron-star
surface. 
To utilize these contours, we assume the properties (equation of state, mass,
radii, etc.) for a 1.358-$M_\odot$ neutron star with $R_{\infty}$ = 15.6~km~--- 
appropriate for models that assume neutron-star cores with moderately stiff
equations of state and containing strong proton superfluidity (e.g., Kaminker,
Haensel, \& Yakovlev 2001; Kaminker, Yakovlev, \& Gnedin 2002 and references
therein). 
For a Crab distance of 2~kpc, the contours (Fig.~\ref{f:bbnorm_kT}) imply a
(2-$\sigma$, 3-$\sigma$) upper limit to the (gravitationally-redshifted)
blackbody surface temperature viewed at infinity of $T_{\infty} < (1.85,
1.97)$~MK [$\log T_{\infty} < (6.27, 6.30)$]. 
This upper limit is slightly less than the one we (Tennant et al.\ 2001)
previously obtained using only the zero-order counting rate~--- no spectral
data~--- from a longer observation.
More difficult to quantify are consequences of departures from uniform,
isotropic, blackbody emission (e.g., Pavlov et al.\ 1994; Zavlin et al.\ 1995;
Pavlov 2000; Becker \& Pavlov 2001).
Hence, we regard $T_{\infty}$ as a representative and indicative upper limit. 

\clearpage

\vspace{10pt}
\begin{center}
\includegraphics[angle=-90,width=\columnwidth]{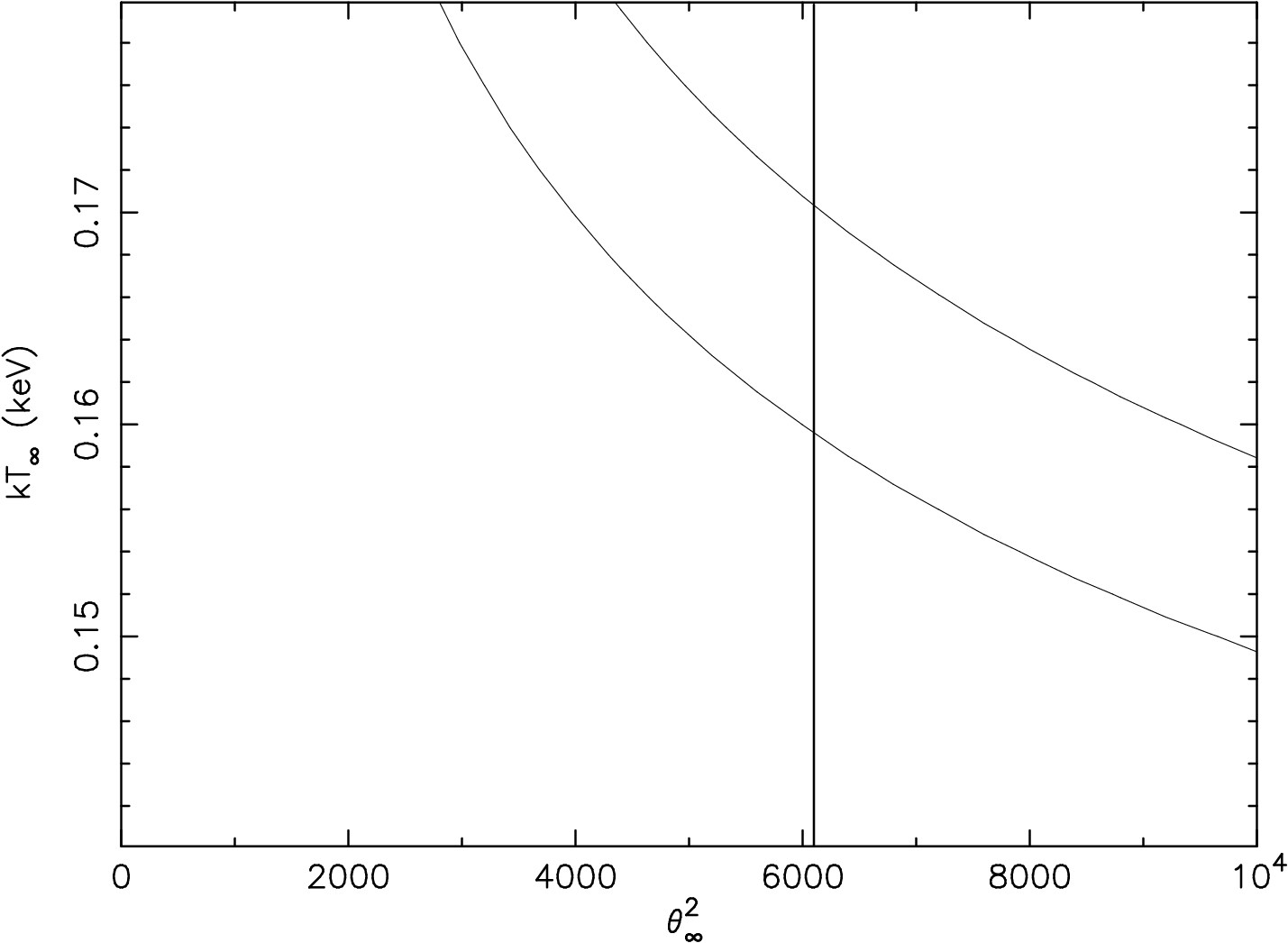}
\vspace{10pt}
\figcaption {Outer confidence contours of the neutron-star surface temperature
versus blackbody normalization. The (lower, upper) contour denotes the the upper 
limit at (95.4\%, 99.7\%) confidence~--- $\Delta\chi^{2} = 
\chi^{2}-\chi^{2}_{\rm min} = (4.0, 9.0)$, corresponding to the single-parameter 
(2-$\sigma$, 3-$\sigma$) upper limit. For reference, our canonical neutron star
has a blackbody normalization $\theta_{\infty}^{2} = 6100\, [{\rm km}/(10\, {\rm 
kpc})]^2$.   \label{f:bbnorm_kT}}
\end{center}
\clearpage

\section{Summary} \label{s:sum}

We have computed and compared the measurements of the phase-averaged spectrum of
the Crab Pulsar using a variety of cross-sections and abundances.
We have shown that results are somewhat sensitive to the abundances and
cross-sections used in the analysis. 
This is especially true for the hydrogen column and emphasizes the importance of
specifying which cross-sections and abundances are assumed in the data
analysis.
We have also compared our results with a number of previous observations.
Although we confirm the results first derived from {\sl Copernicus} data
(Keenan, Hibbets, \& Dufton 1985 and references therein) and more recently of
Willingale et al.\ (2001) that the Crab line-of-sight is underabundant in
oxygen, our analysis of the \cha -LETGS data suggests a somewhat greater
hydrogen column and a smaller (factor of 0.6) relative oxygen abundance
than this previous analysis.
The increased hydrogen column we measure is primarily due to the choice of
cross-sections and abundances, whereas the lower relative oxygen abundance we
attribute to the better spectral resolution and calibration accuracy of the
\cha\ LETGS. 
In addition, we have measured for the first time the spectrum of the Crab Pulsar
as a function of pulse phase at all pulse phases. 
We find marginal evidence for variation of the power law spectral
index, but the statistics at and near pulse minimum are limited. 
Future, more precise, measurments are needed. 
In all our analyses, we have accounted for the contribution of scattering by
interstellar dust to the extinction of X rays in an aperture-limited
measurement~--- a consideration in spectral analysis of point sources observed
with \cha 's exceptional angular resolution. 
Finally, we used the spectral data to obtain a new and better upper limit to the
temperature of the neutron star of $\log T_{\infty} < 6.30$($3-\sigma$).

\acknowledgements {We thank the referee for detailed and insightful
comments.
We also acknowledge a tremendous debt to Leon Van
Speybroeck for his remarkable contributions to the development of the \cha\
optics. His untimely death has touched us all.}

\end{document}